\documentclass[prl,english,twocolumn,preprintnumbers,amsmath,amssymb,nofootinbib,superscriptaddress,longbibliography,noeprint]{revtex4-2}

\pdfoutput=1

\usepackage[latin1]{inputenc}
\usepackage{graphicx}
\usepackage{color}
\usepackage{bbm}
\usepackage{float}
\usepackage{cleveref}
\usepackage{amssymb}
\usepackage{physics}
\usepackage{amsmath}
\usepackage{tabularx}
\usepackage[dvipsnames]{xcolor}

\usepackage{dsfont}

\def\0#1#2{\frac{#1}{#2}}

\def\s0#1#2{\mbox{\small{$ \frac{#1}{#2} $}}}

\newcommand{\beq}{\begin{equation}}
\newcommand{\eeq}{\end{equation}}
\newcommand{\bea}{\begin{eqnarray}}
\newcommand{\eea}{\end{eqnarray}}

\usepackage{babel}
\makeatother
\begin{document}

\title{Fourth- and fifth-order virial expansion of harmonically trapped fermions at unitarity}

\author{Y. Hou}
\affiliation{Department of Physics and Astronomy, University of North Carolina,
  Chapel Hill, North Carolina 27599, USA}

\author{K. J. Morrell}
\affiliation{Department of Physics and Astronomy, University of North Carolina,
  Chapel Hill, North Carolina 27599, USA}

\author{A. J. Czejdo}
\affiliation{Department of Physics and Astronomy, University of North Carolina,
  Chapel Hill, North Carolina 27599, USA}
  
\author{J. E. Drut}
\affiliation{Department of Physics and Astronomy, University of North Carolina,
  Chapel Hill, North Carolina 27599, USA}

\begin{abstract}
By generalizing our automated algebra approach from homogeneous space to harmonically trapped systems,
we have calculated the fourth- and fifth-order virial coefficients of universal spin-$1/2$ fermions in the unitary limit, confined in an isotropic harmonic potential. We present results for said coefficients as a function of trapping frequency 
(or, equivalently, temperature), which compare favorably with previous Monte Carlo calculations (available only at fourth order) as well as with our previous estimates in the untrapped limit (high temperature, low frequency). We use our estimates of the virial expansion, together with resummation techniques, to calculate the compressibility
and spin susceptibility.
\end{abstract}

\maketitle

{\it Introduction.- }
At low enough temperatures, or high enough densities, matter invariably displays
its quantum mechanical nature, first and foremost by virtue of quantum statistics 
(i.e. particles are ultimately bosonic or fermionic, at least in three spatial dimensions) 
but also due to interaction effects that may alter the nature of the equilibrium state.
As the temperature is raised, these systems eventually undergo a quantum-classical
crossover (QCC) in which interactions still play a role, but where quantum mechanical effects 
are slowly washed out by temperature fluctuations.
This regime is especially interesting for strongly coupled matter, in particular
in cases where a superfluid phase is present, as the behavior
above the superfluid critical temperature (i.e. in the un-ordered phase) is still significantly
affected by the interactions (e.g. inducing pairing correlations) but there are no obvious 
effective theory descriptions~\cite{Pieri2004, Sedrakian2005, Lobo2006, Randeria2010, Gaebler2010, Chen2014, Mueller2017, Jensen2019, 
Richie-Halford2020, Rammelmueller2021}.

The QCC is governed by the so-called virial expansion (VE)~\cite{Pathria2011}, which
breaks the quantum many-body problem into $n$-particle subspaces, captured in the so-called virial coefficients (see Ref.~\cite{Liu2013PR} for a review). 
For bulk thermodynamic quantities the virial coefficients are denoted by $b_n$, and their change due to interactions is $\Delta b_n$. The calculation of 
$\Delta b_n$ has a sparse history that started with $\Delta b_2$ in 1937, by Beth and 
Uhlenbeck~\cite{BETH1937915} and remained 
largely quiet until the early 21st century. On the theory side, this quiet period can be attributed to the
well-known fact that the quantum two-body problem is considerably easier to solve (and to relate to two-body scattering properties) than its three- and higher-body counterparts. On the experimental side, these quantum virial coefficients became increasingly relevant in the early 2000's with the rise of ultracold atom experiments around the world and their ever-increasing ability to create, manipulate, and measure atomic clouds~\cite{ExpReview}.

One of the most famous systems studied with ultracold atoms is the so-called unitary
limit of the spin-$1/2$ Fermi gas~\cite{Zwerger2012}, which represents a universal
regime relevant for atomic and nuclear physics~\cite{Ho2004, Braaten2006, RevModPhys.80.885, RevModPhys.80.1215, Levinsen2017JoPBAMaOP, STRINATI20181}.
In this work we investigate the QCC of this universal regime using the VE up to fifth order
for a system confined by a harmonic oscillator (HO) potential.
Previous numerical work calculated 
$\Delta b_3$~\cite{Liu2009PRL,Liu2010PRA,Kaplan2011PRL,Leyronas2011PRA,Gao2015EEL,Endo2016JoPAMaT} and $\Delta b_4$~\cite{Yan2016PRL,Rakshit2012PRA,Ngampruetikorn2015PRA,Endo2015PRA} (see also Ref.~\cite{Gharashi2012PRA,Peng2014PRA, Kristensen2016PRA}).
More recent work~\cite{Morrell2019PRA} studied analytic expressions in the so-called semiclassical approximation (previously implemented in a wide variety of situations~\cite{Shill2018PRA, Hou2019, Berger2020PRA, Czejdo2020, Hou2020PRL,Hou2020PRA}), which uses a coarse discretization of imaginary time.
On the experimental side, there have also been attempts to determine $\Delta b_4$
at unitarity in the untrapped limit, using measurements of the equation of state~\cite{MITExp, ENSExp}, 
However, those analyses are numerically challenging because one must fit a fourth-order polynomial 
assuming higher-order contributions are small (which is not necessarily the case, as
shown in Ref.~\cite{Hou2020PRL}).

In this work we generalize the above calculations to include $\Delta b_5$ and go far beyond the semiclassical approximation, extrapolating to the continuous imaginary time limit. 
While we restrict ourselves to the unitary limit for the most part, 
we provide approximate analytic formulas that apply to arbitrary interaction strengths, 
trap frequency, and spatial dimension.

{\it Hamiltonian and virial expansion.-}
In this work we focus on a system of harmonically trapped spin-$1/2$ fermions interacting via
a short-range interaction. Thus, the Hamiltonian is \( \hat{H} = \hat{T} + \hat{V}_\text{ext} + \hat{V} \), where
\beq
 \hat{T} = \sum_{s = \uparrow,\downarrow} \int d^3 r ~ \hat{\psi}_s^{\dagger} (\mathbf{r}) \left (- \frac{\hbar \nabla^2}{2 m } \right)  \hat{\psi}_s (\mathbf{r}),
\eeq
is the kinetic energy operator,
\beq
\hat V^{}_\text{ext} = \int\! d^3r\, \frac{1}{2} m \omega^2 {\bf r}^2  \left [\hat n_{\uparrow}^{}({\bf r}) + \hat n_{\downarrow}^{}({\bf r}) \right],
\eeq
is the external potential energy operator, and
\beq
\label{Eq:Vint}
\hat{V} = - g \int d^3 r ~ \hat{n}_{\uparrow}(\mathbf{r}) \hat{n}_{\downarrow}(\mathbf{r}),
\eeq
is the interaction.  Above, $m$ is the mass of the particles, 
$\omega$ is the isotropic harmonic trapping frequency, $g$ is the bare coupling,
$\hat n_{s}^{}({\bf r}) = \hat \psi^{\dagger}_s({\bf r}) \hat \psi^{}_s({\bf r})$ is the particle 
density operator for spin-$s$ particles, and $\hat \psi^{\dagger}_s({\bf r})$ and $ \hat \psi^{}_s({\bf r})$ are,
respectively, the creation and annihilation operators for particles of spin $s$ at position $\bf r$. We use units such that $\hbar = k_B =m=1$ from this point on.
Naturally, the noninteracting piece $\hat{T} + \hat V^{}_\text{ext}$ can be diagonalized exactly in the single-particle subspace of the Fock space, 
which leads to the HO basis we will refer to below.
The contact interaction of Eq.~(\ref{Eq:Vint}) is singular in three spatial dimensions and must therefore be regularized and renormalized. To that end, we place the system on a spatial lattice of spacing 
$\ell$ and implicitly take the continuum limit by transforming spatial sums into integrals at 
the end. In the process, we renormalize by tuning the coupling so that the known two-body
answer for the second-order virial coefficient is reproduced (see below).

The VE accesses thermodynamics by breaking down the calculation by particle number. 
Specifically, one expands the grand thermodynamic potential \( \Omega \) in powers of the fugacity \( z = \exp(\beta \mu) \) as
\beq
\label{Eq:OmegaVirialExpansion}
-\beta \Omega = \ln \mathcal{Z} = Q_1 \sum_{n=1}^{\infty} b_n z^n,
\eeq
where $\beta$ is the inverse temperature, $Q_1$ is the single-particle partition function, and \( b_n \) is the \( n \)-th order virial coefficient.
The $b_n$ capture, in a nonperturbative fashion, the contribution of the $n$-body problem to the full $\Omega$. Plugging in the definition of
the grand-canonical partition function \( \mathcal{Z} \), namely
\beq
\mathcal{Z} = \tr\left [e^{-\beta (\hat{H} - \mu \hat{N})} \right] = \sum_{N=0}^{\infty} z^N Q_N,
\eeq
into Eq.~(\ref{Eq:OmegaVirialExpansion}) and expanding $\ln \mathcal{Z}$ in powers of $z$, the $b_n$ can be written in terms of the $N$-particle canonical partition functions
$Q_N = \tr^{}_N \left[ e^{- \beta \hat{H}} \right],$
where the trace is over the \( N \)-particle Hilbert space (see below). 

{\it Computational framework.-}
To evaluate $Q_N$, we implement a symmetric Suzuki-Trotter decomposition 
\beq
 e^{-\beta \hat H} = 
 \lim_{N_{\tau} \to \infty} 
 \left[
 e^{-\frac{\tau}{2} (\hat{T} + \hat V^{}_\text{ext})} e^{-\tau \hat{V}} e^{-\frac{\tau}{2} (\hat{T} + \hat V^{}_\text{ext})}
 \right]^{N_{\tau}},
\eeq
where we split \( \beta = \tau N_{\tau} \) into \( N_{\tau} \) time steps.
Thus,
\beq
  \label{eq:7}
  Q_N = \lim_{N_{\tau} \to \infty} \tr_N 
 \left[
 e^{- \tau (\hat{T} + \hat V^{}_\text{ext})} e^{-\tau \hat{V}} 
\right]^{N_{\tau}},
\eeq
where the cyclic property of the trace was used.
To proceed, we calculate the matrix elements of the factors inside the trace in coordinate space.
For a single imaginary time step, those matrix elements define the factorized transfer matrix $\mathcal M_{ab}$, for $a$ particles of spin-$\uparrow$ and $b$ particles of spin-$\downarrow$. 
For example, in the $1+1$ subspace, i.e. $a = b = 1$, we obtain
\bea
[\mathcal M_{11}]_{{\bf X},{\bf Y}} &=& \langle {\bf X} |  e^{- \tau (\hat{T} + \hat V^{}_\text{ext})} e^{-\tau \hat{V}} | {\bf Y} \rangle \nonumber \\ 
&=&
\rho({\bf x}_1,{\bf y}_1)\rho({\bf x}_2,{\bf y}_2) \left[ \openone + C \delta({\bf y}_1-{\bf y}_2)\right],
\eea
where ${\bf X} = ({\bf x}_1, {\bf x}_2)$, ${\bf Y} = ({\bf y}_1, {\bf y}_2)$, $C = (e^{\tau g/\ell^3} - 1)\ell^3$,
\beq
\rho({\bf x},{\bf y}) = \frac{1}{\lambda^3_T}\left[\frac{\beta \omega}{\sinh(\tau \omega)}\right]^{3/2}\exp[-{\bf Z}^T B {\bf Z}],
\eeq
with $\lambda_T = \sqrt{2 \pi \beta}$, ${\bf Z}^T = ({\bf x}^T/\lambda_T, {\bf y}^T/\lambda_T)$, and
\bea
B = 
\frac{\pi \beta \omega}{\sinh(\tau \omega)}
\left(
\begin{array}{cc}
\cosh(\tau \omega)\openone & -\openone\\
-\openone  & \cosh(\tau \omega)\openone
\end{array}
\right),
\eea
where $\openone$ is a $3\times3$ unit matrix.

While the above example does not involve identical particles, for the
cases that do (e.g. the $2+1$ subspace of the 3 particle Hilbert space), the (anti-)symmetrization 
can be carried out at the very end, i.e. after taking the $N_\tau$-th power of the
distinguishable-particle transfer matrix. This property was already noted by Huang and Yang
in 1959~\cite{PhysRev.113.1165, PhysRev.116.25} and is a consequence of the fact that the operators involved do not change the
particles' statistics. Thus, there is no need to use (anti-)symmetrized intermediate states in the
calculation, which greatly reduces the computational effort.
In the Supplemental Materials we report on the generalization of the above result to 
$\mathcal M_{21}$, $\mathcal M_{31}$, $\mathcal M_{22}$, $\mathcal M_{41}$, and $\mathcal M_{32}$.

Armed with the above factorized transfer matrices $\mathcal  M_{ab}$, we use automated algebra to
symbolically expand $\left[ \mathcal M_{ab}\right]^{N_\tau}$ for varying $N_\tau$.
We then combine the results to obtain the relevant $Q_N$ and from them the $b_n$,
which are in turn extrapolated to the large-$N_\tau$ limit.
More explicitly, the interaction-induced change $\Delta b_n$, for $n=2,3,4,5$ is calculated as
$\Delta b_2 = \Delta b_{11}$, $\Delta b_3 = 2\Delta b_{21}$, $\Delta b_4 = 2\Delta b_{31} + \Delta b_{22} $, and $\Delta b_5 = 2\Delta b_{41} + 2\Delta b_{32}$,
where the subspace contributions are
\bea
\Delta b_{11} &=& {\Delta Q_{11}}/{Q_1}, \\
\Delta b_{21} &=& {\Delta Q_{21}}/{Q_1}  - {\Delta Q_{11}}/{2},
\eea
and the 4- and 5-particle subspaces are shown in the Supplemental Materials.
Here, $\Delta X$ represents the change in $X$ induced by the interactions and the 
$Q_{ab}$ are the canonical partition functions for $a$ particles of spin-$\uparrow$ and $b$ 
particles of spin-$\downarrow$. In the above expressions, the $\Delta b_{ab}$
are intensive quantities, whereas the $Q_{ab}$ themselves scale as $V^{a+b}$ where
$V$ is the spatial volume. That property emphasizes the challenge in calculating $\Delta b_{ab}$
numerically: the delicate cancellations must be resolved among the various terms involving different $Q_{ab}$'s.
It is for that reason that automated algebra methods are advocated here, where those cancellations
can be resolved using arbitrary precision arithmetic, avoiding stochastic effects.

{\it Results: Approximate analytic expressions for $\Delta b_n$.-}
For $N_\tau = 1,2$, we carry out calculations entirely analytically
in which the coupling strength, the trapping frequency, and the spatial dimension
appear as arbitrary variables (in principle, it is also possible to take this to even higher order, 
but the formulas become extremely long). The resulting formulas for $\Delta b_3$ and $\Delta b_4$
for $N_\tau = 1$, first shown in Ref.~\cite{Morrell2019PRA}, qualitatively 
(and in some parameter regions quantitatively) capture the behavior of $\Delta b_n$. 
These formulas are also useful as checks for codes that implement higher values of $N_\tau$.
Here, we provide results broken down by subspace for up to five particles, shown in full detail in the 
Supplemental Materials. 
%

\begin{figure}[t]
  \centering
  \includegraphics[width=1.02\linewidth]{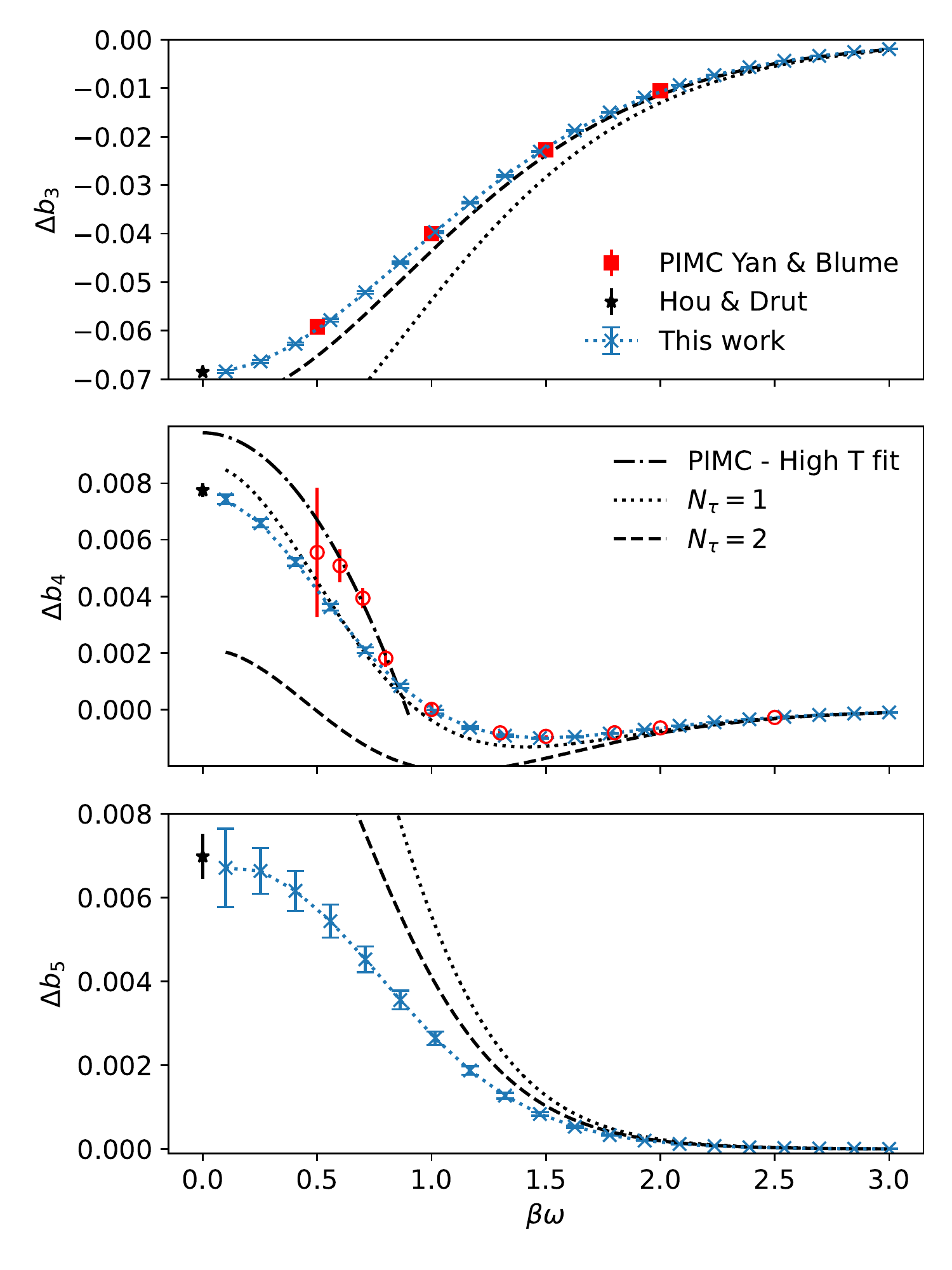}    
  \caption{\( \Delta b_3 \) (top), \( \Delta b_4 \) (middle), and \( \Delta b_5 \) (bottom) as functions of \( \beta \omega \), for a trapped unitary Fermi gas. Our results are shown with blue crosses and error bars, joined by a blue dotted line. The data by Yan and Blume from Ref.~\cite{Yan2016PRL} appears as red squares for $\Delta b_3$ and red circles for $\Delta b_4$, in both cases with error bars.
  The dashed-dotted line in the middle plot shows a high-temperature fit to the data of Ref.~\cite{Yan2016PRL}.
  Black stars with error bars show the results by Hou and Drut from Ref.~\cite{Hou2020PRL} calculated in the homogeneous gas limit. The dotted (dashed) line shows the $N_\tau = 1$ ($N_\tau = 2$) results given analytically in the Supplemental Materials. The latter show that, for \( \Delta b_3 \), 
  increasing $N_\tau$ from 1 to 2 shows a dramatic improvement, whereas the case of \( \Delta b_4 \) 
  is a cautionary tale: as $N_\tau$ goes from 1 to 2, the results move away from our final answer
  (blue crosses). In fact, it is not until $N_\tau = 5$ that \( \Delta b_4 \) reaches the
  asymptotic regime one can use for extrapolation. Reference~\cite{Rakshit2012PRA} presented a large-$\beta\omega$ asymptotic formula
  for $\Delta b_n$, but its validity is well outside the $0 < \beta\omega < 3$ region studied here.
  }
  \label{fig:db3db4db5-betaomega}
\end{figure}

From those formulas we learn that (as shown in Fig.~\ref{fig:db3db4db5-betaomega}), 
increasing $N_\tau$ does not immediately improve the quality of the final answer; rather, the results 
could move {\it away} from the $N_\tau \to \infty$ limit 
before the asymptotic regime is reached, usually for $N_\tau > 2$. Simply put, as $N_\tau$ is increased the 
results may worsen before they improve.
Thus, it is important to investigate as large $N_\tau$ as possible, even if low values are qualitatively 
correct. In our automated calculations, we explored up to $N_\tau = 20$ (for 
$\Delta b_{21}$), $16$ ($\Delta b_{31}$), $12$ ($\Delta b_{22}$), $12$ ($\Delta b_{41}$), 
and $8$ ($\Delta b_{32}$), which we used to estimate the full $\Delta b_3$, 
$\Delta b_4$, and $\Delta b_5$, extrapolated to $N_\tau \to \infty$. 

{\it Results: Virial coefficients in the unitary limit.-}
In our approach, we calculate $\Delta b_2$ as a function of the bare coupling $C$, $\beta\omega$, 
and $N_\tau$, and renormalize by tuning $C$ to the known result in the unitary 
limit~\cite{Liu2013PR}, namely
$\Delta b_2 = \left[ 4 \cosh(\beta \omega/2) \right]^{-1}.$
Thus, the second-order VE is reproduced exactly by virtue of this renormalization
condition, such that the line of constant physics is followed in the extrapolation to 
$N_\tau \to \infty$, for each $\beta \omega$ (see Supplemental Materials of
Ref.~\cite{Hou2020PRL}).

In Fig.~\ref{fig:db3db4db5-betaomega} we show our results for $\Delta b_{3}$ (top),
$\Delta b_{4}$ (center), and $\Delta b_{5}$ (bottom) for a unitary Fermi gas in a harmonic trap
as a function of $\beta \omega$. The error bars represent the uncertainty in the
$N_\tau \to \infty$ extrapolation, given by the difference between the maximum and minimum
predictions of polynomial extrapolation schemes (degrees 2 to 5 for $\Delta b_3$ and
$\Delta b_4$, and degrees 2 and 3 for $\Delta b_5$, where the data is more limited; see
Ref.~\cite{Hou2020PRL}).  Our results for $\Delta b_{3}$ are in superb agreement with the
quantum Monte Carlo data of Ref.~\cite{Yan2016PRL} as well as with the
homogeneous-limit answer of Ref.~\cite{Hou2020PRL}.
[The homogeneous limit is related to the results shown here by \( \Delta b_n^{\mathrm{h}} = n^{3/2} \Delta b_n (\beta \omega \to 0)\),
see Refs.~\cite{McCabe1991PLB, Liu2013PR, Liu2009PRL, Liu2010PRA}]. The case of
$\Delta b_{4}$ is less clear cut: there is good agreement with Ref.~\cite{Yan2016PRL} for
$\beta \omega \geq 1$, but a clear difference remains at low frequencies. We return to this issue
below.  Finally, we predict $\Delta b_{5}$ as a function of $\beta \omega$, which to the best of
our knowledge does not appear elsewhere in the literature.  As the HO potential
confines the system, it naturally increases its kinetic energy, effectively reducing
the interaction effects. This suggests that, for a given interaction strength, the VE
should enjoy better convergence properties when a trapping potential is turned on (as
argued also in Ref.~\cite{Liu2013PR}).  Indeed, although our results indicate that
$\Delta b_4 \simeq \Delta b_5$ and, moreover, for $0.3 < \beta \omega < 1.4$ we find
$\Delta b_5 > |\Delta b_4|$, we also find that $|\Delta b_2| \gg |\Delta b_3| \gg |\Delta b_4|$.
\begin{figure}[t]
  \centering
  \includegraphics[width=\linewidth]{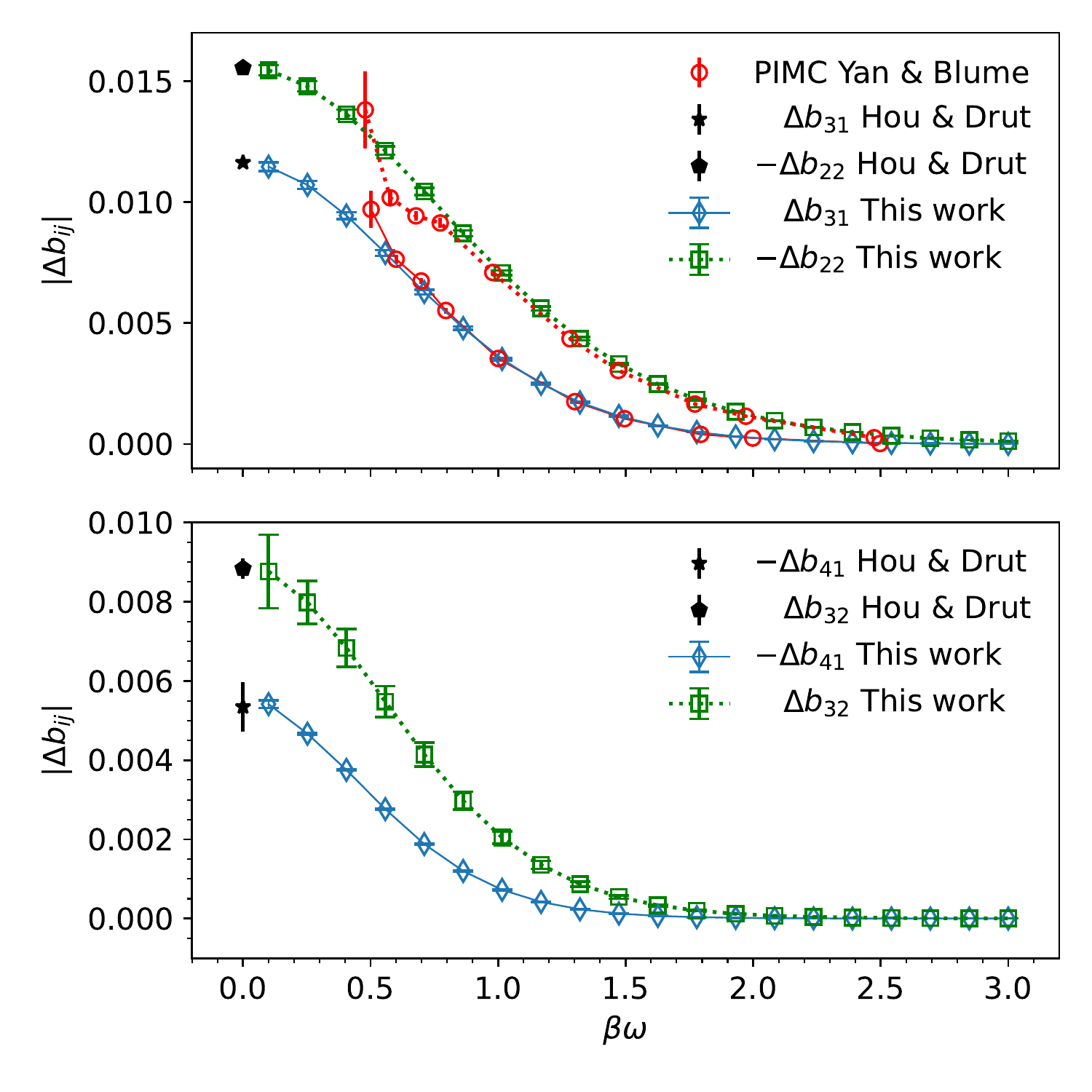}      
   \caption{{\bf Top:} \( \Delta b_{31} \) (blue diamonds) and \( -\Delta b_{22} \) (green squares) as functions of \( \beta \omega \), compared with the PIMC results of Ref.~\cite{Yan2016PRL} (red circles joined by solid line for \( \Delta b_{31} \) and joined by a dotted line for \( \Delta b_{22} \)).
   The black star and black pentagon show, respectively, the results for \( \Delta b_{31} \) and \( -\Delta b_{22} \) in the $\beta \omega \to 0$ limit, obtained in Ref.~\cite{Hou2020PRL}.
   {\bf Bottom:} \(-\Delta b_{41} \) (blue diamonds) and \( \Delta b_{32} \) (green squares) as functions of \( \beta \omega \).
   The black star and black pentagon show, respectively, the results for \( -\Delta b_{41} \) and \( \Delta b_{32} \) at $\beta \omega = 0$ from Ref.~\cite{Hou2020PRL}.
   }
  \label{fig:db31db22-betaomega}
\end{figure}

To better understand the differences in $\Delta b_4$ between our results and
Ref.~\cite{Yan2016PRL}, we plot in Fig.~\ref{fig:db31db22-betaomega} 
(top panel) the subspace contributions $\Delta b_{31}$ and $\Delta b_{22}$. As pointed 
out in Ref.~\cite{Yan2016PRL}, these contributions partially cancel
each other out, leading to the observed increased uncertainty 
in the final answer. Clearly, the largest differences arise in the determination 
of $\Delta b_{22}$, which is not unexpected as a contact interaction in that subspace is less
susceptible to Pauli blocking than $\Delta b_{31}$.

Figure~\labelcref{fig:db31db22-betaomega} (bottom panel) shows our results for \( \Delta b_{41} \) 
and \( \Delta b_{32} \), whose behavior parallels  \( \Delta b_{31} \) 
and \( \Delta b_{22} \) in that they enter with different signs but similar magnitude,
thus leading to increased uncertainty in the final result for $\Delta b_5$. 
In spite of those delicate cancellations, we are able to resolve the fifth-order
contribution, as shown already in the bottom panel of Fig.~\ref{fig:db3db4db5-betaomega}.
Nevertheless, the size of the error bars of $\Delta b_{32}$ is larger than that of
 $\Delta b_{41}$. This may come as a surprise
 given the results of Ref.~\cite{Hou2020PRL},
 whose uncertainty at $\beta \omega = 0$ for $\Delta b_{41}$ is larger than for $\Delta b_{32}$.
Those results were calculated at the same $N_\tau=9$ order for both coefficients, 
using an analytic cancellation of volume-dependent terms.
In contrast, in the present work we achieved $N_\tau=12$ for $\Delta b_{41}$ but only $N_\tau = 8$ for 
 $\Delta b_{32}$, due to the increasing computational cost of cancelling the volume-dependent terms, 
which is done numerically in the trapped case.

{\it Results: Applications to thermodynamics.-}
Having obtained the precise form of $\Delta b_3$, $\Delta b_4$, and $\Delta b_5$ 
as functions of $\beta \omega$ for harmonically trapped fermions in the unitary limit, 
we apply those results to obtain thermodynamic information. As an example, we report here
the compressibility and magnetic susceptibility, respectively $\chi^{}_n$ and $\chi^{}_s$,
defined as $\chi^{}_{n,s} = \beta^{-1} \partial^2 \ln \mathcal Z / \partial h_{\pm}^2$,
where $h^{}_{\pm}  = (\mu_\uparrow \pm \mu_\downarrow)/2$ and $\mu_s$ is the chemical 
potential for spin-$s$ particles. The interaction effects on $\chi_{s,n}$ are
\beq
\Delta \chi_{n,s} = \frac{\lambda_T^2}{8 \pi} Q_1 \sum_{n=3}^{\infty} \sum_{m+j=n} (m \pm j)^2 \Delta b_{mj} z_{\uparrow}^m z_{\downarrow}^j,
\eeq
where $z_s = e^{\beta \mu_s}$ is the fugacity for spin-$s$ particles. 
\begin{figure}[t]
  \centering
  \includegraphics[width=\linewidth]{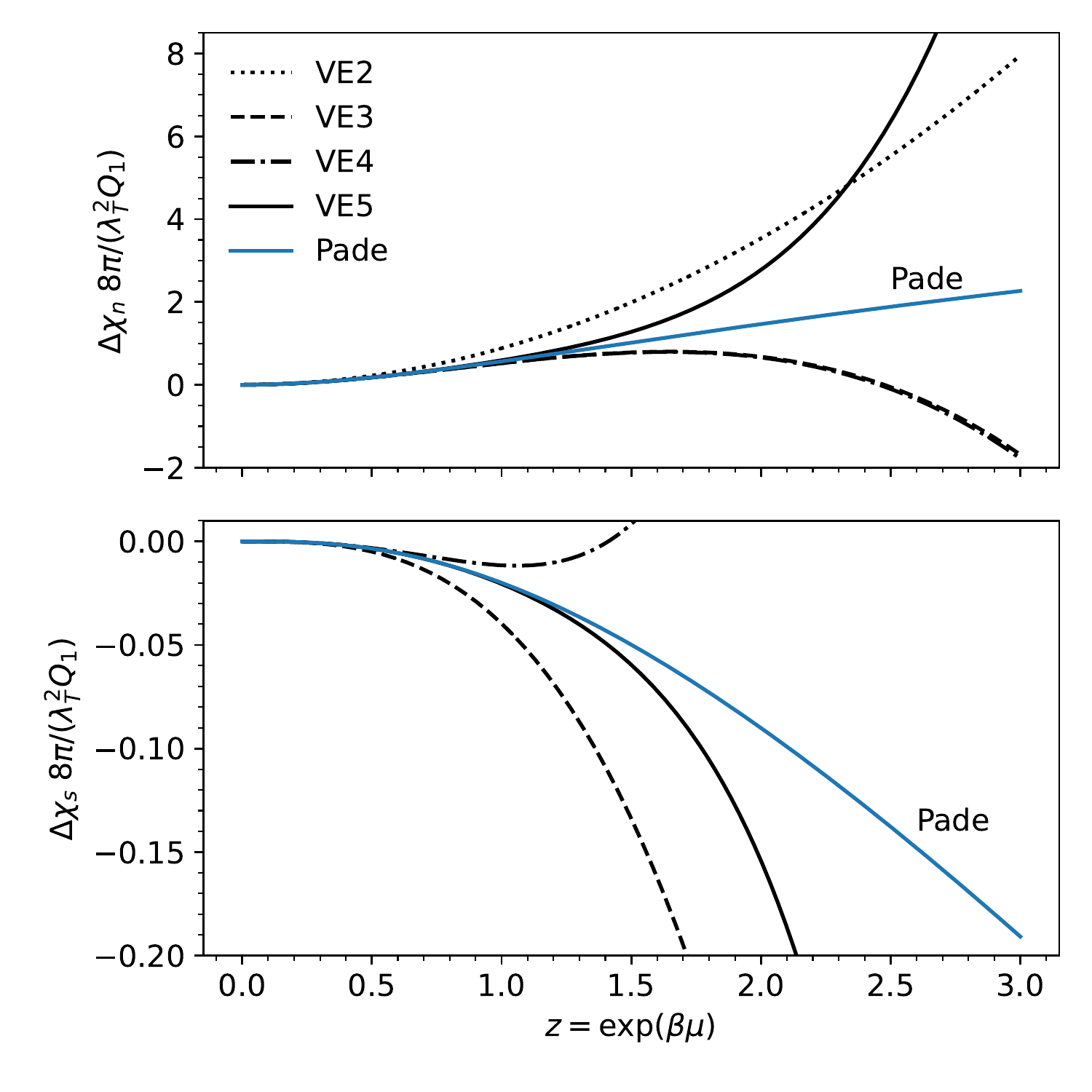}      
   \caption{{\bf Top}: Interaction effects on the compressibility $\Delta \chi^{}_n$, in 
   units of $8\pi / (\lambda_T^2 Q_1)$, as a function of the fugacity $z$ for
   a harmonically trapped unitary Fermi gas at $\beta \omega = 1$. The second, third, fourth, 
   and fifth-order VE results are shown, respectively, with dotted, dashed, dash-dotted,
   and solid lines. The Pad\'e resummed result (with a $[3/2]$ approximant) is shown
   as a blue line.
   {\bf Bottom}: Interaction effect on the magnetic susceptibility $\Delta \chi^{}_s$ as a function of 
   $z$, for the same parameters as in the top panel. The second-order VE is omitted because 
   it is identically zero for $\Delta \chi^{}_s$.
   }
  \label{fig:kappa-chi}
\end{figure}
Our results, shown in Fig.~\ref{fig:kappa-chi}, indicate that the partial sums of the
VE display large variations for $\Delta \chi^{}_{n,s}$ as the VE order is increased, in
particular for $z \ge 1$. However, we also see that, using the high-order coefficients
we calculated here, it is possible to carry out a Pad\'e resummation [and related strategies (see e.g.~\cite{Rossi2018PRL})] to
obtain sensible results for static response functions even as far as $z = 3$.

{\it Conclusion and outlook.-}
In this work we have determined the frequency dependence of 
the virial coefficients $b_n$ of HO-trapped spin-$1/2$ fermions at
unitarity. We used a discretization of the imaginary time direction and a 
Suzuki-Trotter factorization of the transfer matrix, together with automated algebra 
methods, to calculate canonical partition functions and from them the interaction 
induced change $\Delta b_n$, for $n = 3,4,5$, which we extrapolated to the
continuous-time limit. To complement those numerical results, we provided
analytic formulas for $\Delta b_n$ in coarse lattices for arbitrary trap frequency and 
spatial dimension.
Using our final $\Delta b_n$, we calculate the compressibility and susceptibility
of the unitary Fermi gas and showed that the VE can be Pad\'e-resummed to obtain
sensible results even as far as $z=3$.

\clearpage
\acknowledgments
This material is based upon work supported by the
National Science Foundation under Grants No.~PHY{1452635}
and No.~PHY{2013078}.

\bibliographystyle{apsrev4-2}
\bibliography{TrappedVirialRefs}

\end{document}